\newcommand{\ud}{\mathrm{d}}
\newcommand{\Graph}[2][0.3]{\vcenter{\hbox{\includegraphics[scale=#1]{#2}}}}

\documentclass{PoS}
\usepackage[utf8]{inputenc}
\usepackage{amsmath,amssymb,amscd,url}

\title{Mixed QCD-EW two-loop corrections to Drell-Yan production}
\ShortTitle{Mixed QCD-EW two-loop corrections to Drell-Yan production}
\author{\speaker{Matthias Heller}\\
        PRISMA$^+$ Cluster of Excellence, \\
        Institut f\"ur Kernphysik, Johannes Gutenberg Universit\"at,\\ 55116 Mainz, Germany\\
        E-mail: \email{maheller@students.uni-mainz.de}}

\author{Andreas von Manteuffel\\
        Department of Physics and Astronomy, Michigan State University, \\
East Lansing, Michigan 48824, USA\\
        E-mail: \email{vmante@msu.edu}}
        
\author{Robert M. Schabinger\\
        Department of Physics and Astronomy, Michigan State University, \\
East Lansing, Michigan 48824, USA\\
        E-mail: \email{schabing@msu.edu}}
        
        \author{Hubert Spiesberger\\
        PRISMA$^+$ Cluster of Excellence, \\
        Institut f\"ur Physik, Johannes Gutenberg Universit\"at,\\ 55116 Mainz, Germany\\
        E-mail: \email{spiesber@uni-mainz.de}}

\abstract{The Drell-Yan production of charged lepton pairs is one of the key processes measured at hadron colliders. The QCD corrections to the cross section are known to order $\alpha_s^2$ and electroweak corrections are known to order $\alpha$. The next important step for a better theoretical understanding is the complete calculation of the mixed QCD-EW corrections of order $\alpha_s\alpha$. In my talk, I report on the first complete calculation of the virtual two-loop corrections of order $\alpha\alpha_s$ to the lepton-pair production cross section. The calculation is carried out analytically using tensor reduction, integration-by-parts relations and the method of differential equations. We validate a previous calculation of the subset of mixed QCD-QED corrections and show how the jet and soft functions of that reference can be used to subtract the infrared divergences of the complete mixed QCD-electroweak virtual corrections.}

\FullConference{14th International Symposium on Radiative Corrections (RADCOR2019)\\ 
9-13 September 2019\\
		Palais des Papes, Avignon, France}

\begin{document}

\section{Introduction}
The Drell-Yan process~\cite{Drell:1970wh} is one of the key processes at the Large Hadron Collider at CERN. The measurement of its cross section can be used to determine the masses of the $W^{\pm}$ and $Z$ bosons, the weak mixing angle, as well as parton distribution functions. In order to have an accurate understanding and better prediction of the process from the theory side, higher-order corrections must be included. Known corrections to the cross section include up to now: next-to-next-to-leading order (NNLO) Quantum Chromodynamic (QCD) corrections \cite{Hamberg:1990np,Harlander:2002wh,Anastasiou:2003ds,Melnikov:2006di}, NNLO Quantum Electrodynamics (QED) corrections \cite{Berends:1987ab,Blumlein:2019srk} and NLO  electroweak (EW) corrections \cite{Baur:2001ze,Dittmaier:2001ay,Baur:2004ig}. Since these corrections turn out to be important, it is natural to also consider the two-loop mixed QCD-EW corrections. At present, results are known for the mixed QCD-QED corrections \cite{Kilgore:2011pa} and in the approximation of on-shell $Z$ production \cite{Dittmaier:2014qza,Dittmaier:2015rxo,Delto:2019ewv,Bonciani:2019nuy}.

In this talk at RADCOR2019, I report on the first calculation of all virtual mixed QCD-EW two-loop corrections. We calculated the master integrals contributing to this process for the first time in the physical region of phase space in terms of multiple polylogarithms and demonstrated that this is possible in the presence of algebraic letters involving unrationalizable square roots \cite{Heller:2019gkq}. For the amplitude we find  exactly the same IR structure as in the case of QCD-QED, such that soft- and jet-functions from that calculation can also be used in the extension to the full EW sector.

\section{Calculation of the amplitude}
\begin{figure}
\centering
\includegraphics[trim={3cm 22cm 2cm 2cm},clip]{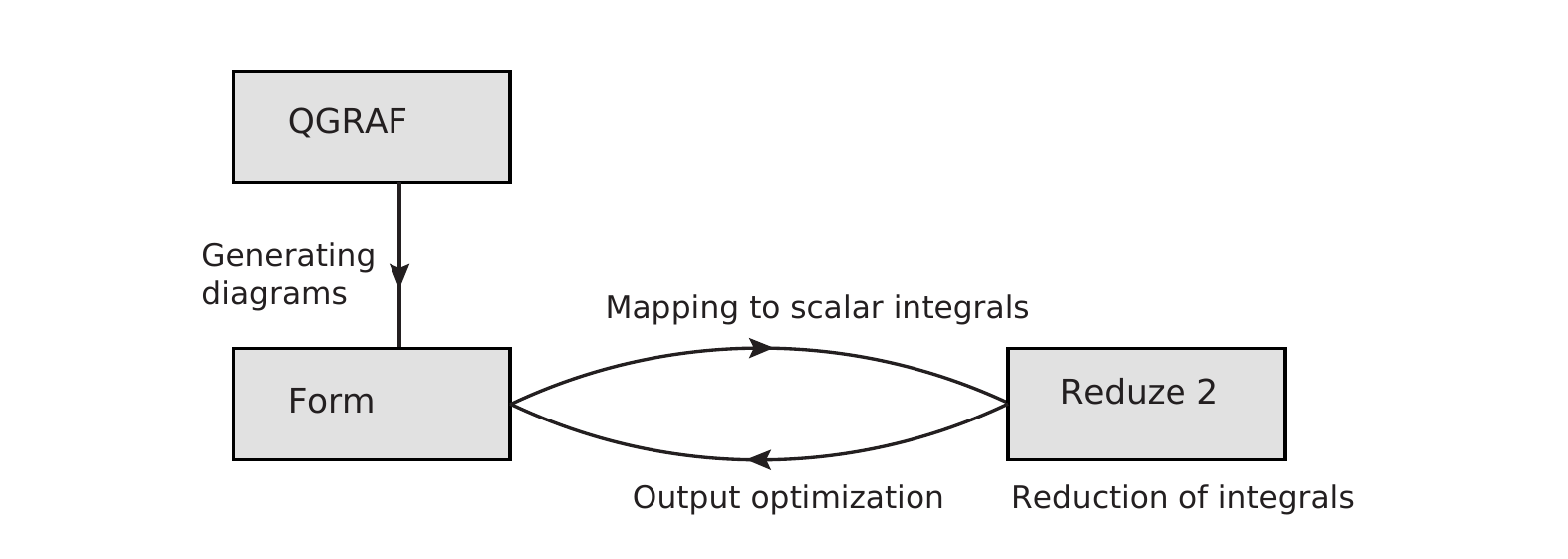} 
\caption{Calculation flow
}
\label{fig:calcflow}
\end{figure}

In this talk, I focus on the mixed QCD-EW two-loop corrections to charged leptons in quark-antiquark annihilation,
\begin{equation}
    q\bar{q} \to l^+ l^-,
\end{equation}
where $l$ is a massless electron or muon.
For the calculation of the amplitude, we use the program \texttt{QGRAF} \cite{Nogueira:1991ex} to generate all contributing diagrams, \texttt{Form} \cite{Ruijl:2017dtg} to apply Feynman rules and do symbolic manipulations, and \texttt{Reduze\;2} \cite{vonManteuffel:2012np,Studerus:2009ye,Bauer:2000cp,fermat} to generate integration-by-parts (IBP) relations. The bare amplitude is written as a sum of a set of master integrals with rational functions in the kinematic variables and the space-time dimension $d$ as coefficients. We use dimensional regularization with $d=4-2\epsilon$ for the regularization of infrared (IR) and ultraviolet (UV) singularities. All rational functions are written in a unique way using partial fractioning. In Fig.~\ref{fig:calcflow}, we show a flow chart of the calculation process.

To evaluate the master integrals, we use the method of differential equations. Using a specific choice for the basis of master integrals, the differential equation can be cast in the so-called canonical form, in which the dependence on $\epsilon$ decouples from the dependence of the kinematic invariants, in such a way
that \cite{Kotikov:2010gf,Henn:2013pwa,Remiddi:2017har,Adams:2018yfj}:
\begin{equation}\label{EqDENormal}
\ud\mathbf{m}_i = \epsilon\; \sum_{j,k} \ud \ln(l_k) \big(A^{(k)}\big)_{ij} \, \mathbf{m}_j,
\end{equation}
where $\mathbf{m}_i$ is a master integral, $\big(A^{(k)}\big)_{ij}$ is an element of a rational matrix and only the $l_k$ (the so-called letters of the differential equation) depend on the kinematic invariants.
The master integrals relevant here have been solved in the Euclidean region of the phase space in Ref.~\cite{Bonciani:2016ypc} and for the case of one internal mass in the physical region of phase space in Ref.~\cite{vonManteuffel:2017myy}.
Here, we aim at completing the analytical calculation of all integrals in terms of functions, which permit a fast and robust numerical evaluation in the physical region of phase space.

\section{Integrating root valued symbols}\label{masterInt}

For the integrals involving two internal masses we find that some of the letters of the differential equation involve unrationalizable square roots. In the literature, such cases were up to now always solved in terms of Chen iterated integrals, in which one integration has to be performed numerically.
In~\cite{Heller:2019gkq}, we were able to integrate the differential equation for the first time in terms of generalized polylogarithms in that region of phase space, rendering the numerical evaluation of the amplitude accessible and efficient for practical applications.

The square roots enter the differential equation the first time as leading singularities in the following three integrals:
\begin{equation}
\epsilon^2 s r_1 \Graph[.34]{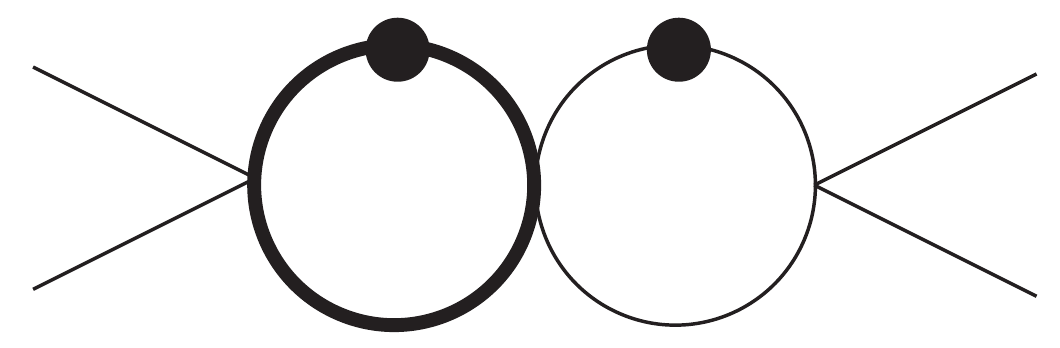},\qquad\epsilon^3 r_2 \Graph[.4]{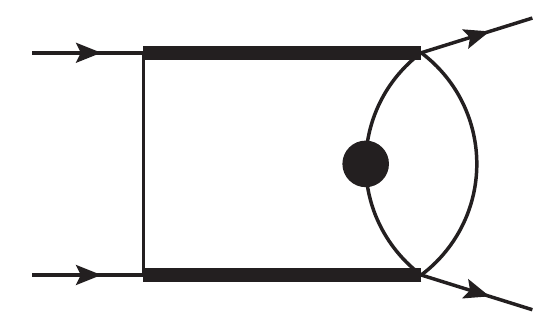},\qquad\epsilon^4 r_3 \Graph[.4]{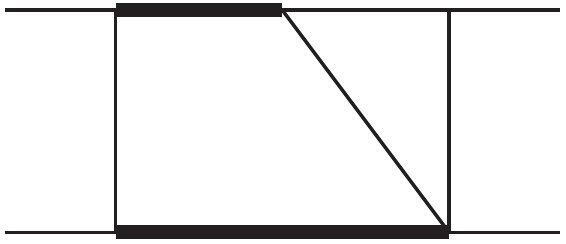},
\end{equation}
where
\begin{align}
r_1&=\sqrt{s(s-4m^2)},\qquad r_2=\sqrt{-st(4m^2(t+m^2)-st)},\nonumber\\
r_3&=\sqrt{s(t^2(s-4m^2)+sm^2(m^2-2t))}.
\end{align}
Following Ref.~\cite{Bonciani:2016ypc}, one can define the dimensionless parameters $w$ and $z$ through:
\begin{align*}
s&=-m^2\frac{(1-w)^2}{w},\qquad t=-m^2\frac{w(1+z)^2}{z(1+w)^2}
\end{align*}
such that two square roots become rational:
\begin{align}
r_1&=\frac{-m^2(1-w)(1+w)}{w},\qquad r_2=\frac{-m^4(1-w)(1-z)(1+z)}{z(1+w)}.
\end{align}
However, the third square root becomes:
\begin{align}
r_3=\frac{m^4(1-w)}{wz(1+w)}r,
\end{align}
where $r$ is a new square root in $w$-$z$-space:
\begin{equation}
r=\sqrt{(1+w^2z^2)(w+z)^2+2wz(w-z)^2+4wz^2(1+w^2)}.
\end{equation}
It can be shown that no parametrization exists which makes $r$ rational \cite{Besier:2019hqd}. Therefore, to integrate the differential equation new methods are needed. 

Using the method of Ref.~\cite{Duhr:2011zq}, one can try to match the symbol defined by the differential equation to a space of functions. In Ref.~\cite{Duhr:2011zq}, this is discussed for rational functions. The idea is to construct suitable ${\rm Li}$ function arguments, such that the functional basis contains no spurious letters.  By considering the symbol of ${\rm Li}$ functions
\begin{equation}
S\big({\rm Li}_n(f)\big)=-(1-f)\otimes \underbrace{f\otimes...\otimes f}_{(n-1)~\text{times}},
\end{equation}
it is clear, that we can achieve the absence of spurious letters by requiring both $f$ and $1-f$ to factorize over the alphabet. A similar strategy can be applied to multiple polylogarithms of several arguments.

In the presence of algebraic letters, we use a heuristic test for factorization. Specifically, we want to test if a given expression $g$ factorizes over the alphabet, i.e.\ can be written as a power-product of the letters:
\begin{equation}
\label{powerproduct}
g = c^{a_0} l_1^{a_1} l_2^{a_2} \cdots,
\end{equation}
where  $c$ and $a_n$ are rational numbers. This implies
\begin{equation}
\label{logrelations}
\ln(g) - a_0 \ln(c) - a_1 \ln(l_1) - a_2 \ln(l_2) - \ldots = 0\,.
\end{equation}
Eq.~\eqref{logrelations} can then be tested numerically to find the required relations.

However, there are several problems with this approach:
\begin{itemize}
\item There is no unique factorization in the case of algebraic letters; e.g.\ if one has letters $\sqrt{x}$ and $\sqrt{y}$, it is unclear, if we need to factor $x-y=(\sqrt{x}-\sqrt{y})(\sqrt{x}+\sqrt{y})$.
\item One needs to consider non-integer powers of letters, e.g.\ $\sqrt{l}$, $l^{1/4},...$. 
\item Often one finds very complicated letters and relations.  
\item There is no obvious way how to choose letters.
\end{itemize}
These problems can be tackled based an interesting observation: we found that, for each $l=q_1+q_2 r$, where $q_1$ and $q_2$ are rational functions,  $\bar{l}=q_1-q_2 r$ is always factorizable over the alphabet, such that one can trade $l$ for $\bar{l}$ without changing the singularity structure. This has a very useful consequence: one can try to get a much simplified version of the alphabet, by making the ansatz
\begin{equation}
l=q+r,\quad \bar{l}=q-r
\end{equation}
and requiring that $l\bar{l}$ factorizes over the rational part of the alphabet. In this way, one can construct a set of simpler letters and construct the algebraic part of the alphabet in a systematic way.

To demonstrate the power of this approach, we show the starting point for the alphabet in the Drell-Yan case. The rational part of the alphabet is given by
\begin{align}
\mathcal{L}_r=\{&1-w,-w,1+w,1-w+w^2,1-z,-z,1+z\nonumber\\
&1-wz,1+w^2z,-z-w^2,z-w\} .
\end{align}
Before simplification, the algebraic part reads
\begin{align}
    \mathcal{L}_a=\{&r, -(1 - w)(z - w)(1 - w\,z)+r\,(1 + w),-(1 - w)\left(\vphantom{w^2}4 w\,z+(w + z) (1 + w\,z)\right)\nonumber\\& -r\,(1 + w), r^2-2 w\,z^2(1 - w)^2 + r\,(w + z) (1 + w\,z),
    \nonumber\\
    &r^2 (1 - z)^2 + 2 z^2 (z + w^2) (1 + w^2 z) + r\,(1 - z)(1 + z)\left(\vphantom{w^2}2 w\,z - (w + z) (1 + w\,z)\right)\} 
\end{align}
and the highest degree of a letter is therefore $8$. Using the ideas described above, we find a simplified version of the algebraic part,
\begin{align}
    \tilde{\mathcal{L}}_a &=\{r, \frac{1}{2}\big(2 + z - w + w\,z (w + z) + r\big), \frac{1}{2}\big(2 w^2 + z - w + w\,z (w + z) + r\big), \nonumber\\
    &\qquad\frac{1}{2}\big(-(w + z)(1 - w\,z)+r\big), \frac{1}{2}\big(-(z - w)(1 + w\,z)+r\big)\},\label{letterssimple}
\end{align}
where the highest degree in $w$ and $z$ is now only $3$. Note, that the factors of $1/2$ in Eq.~\eqref{letterssimple} are chosen to avoid having to introduce 2 as an auxiliary letter. With the new representation of the alphabet, we were able to integrate the differential equation in the physical region of phase space, using only ${\rm Li}_3$ and ${\rm Li}_{21}$ functions for weight $3$ and ${\rm Li}_4$, ${\rm Li}_{22}$ and ${\rm Li}_{31}$ functions for weight 4. We were able to choose a representation in which all functions are manifestly real-valued above the two-mass threshold.

\section{UV renormalization and IR structure}
\subsection{The case of QCD-QED}
In Ref.~\cite{Kilgore:2011pa} the authors calculated the mixed QCD-QED corrections to lepton-pair production. The UV renormalization in this case is trivial, since it only affects the vacuum polarization diagrams and does not mix with IR divergences. To subtract IR divergences, the authors calculated jet and soft functions, which are given by
\begin{align}
\mathcal{J}^{(1,0)}&=-\left(\frac{1}{2\epsilon^2}+\frac{3}{4\epsilon}\right)(Q_q^2+Q_l^2),\qquad\mathcal{J}^{(0,1)}=-\left(\frac{1}{2\epsilon^2}+\frac{3}{4\epsilon}\right)C_F,\nonumber\\
\mathcal{J}^{(1,1)}&=\left(\frac{1}{4\epsilon^4}+\frac{3}{4\epsilon^3}+\frac{9}{16\epsilon^2}\right)C_F(Q_q^2+Q_l^2)-\frac{1}{2\epsilon}\left(\frac{3}{16}-\frac{3}{2}\zeta_2+3\zeta_3\right)C_FQ_q^2,\nonumber\\
\mathcal{S}^{(1,0)}&=-\frac{1}{2\epsilon}\left[(Q_q^2+Q_l^2)\ln\left(\frac{\mu^2}{-s}\right)+2Q_qQ_l\left(\ln\left(\frac{\mu^2}{-t}\right)-\ln\left(\frac{\mu^2}{-u}\right)\right)\right],\nonumber\\
\mathcal{S}^{(0,1)}&=-\frac{1}{2\epsilon}C_F\ln\left(\frac{\mu^2}{-s}\right),\nonumber\\
\mathcal{S}^{(1,1)}&=\frac{1}{4\epsilon^2}C_F\ln\left(\frac{\mu^2}{-s}\right)\left[(Q_q^2+Q_l^2)\ln\left(\frac{\mu^2}{-s}\right)+2Q_qQ_l\left(\ln\left(\frac{\mu^2}{-t}\right)-\ln\left(\frac{\mu^2}{-u}\right)\right)\right],\label{SoftJet}
\end{align}
where $Q_q$ is the charge of the quark, $Q_l$ the charge of the lepton and $C_F$ is the color factor of $SU(N_c)$.
One can then extract the finite hard scattering amplitude $\mathcal{H}$ from the bare amplitude $\mathcal{M}$ by:
\begin{align}
\mathcal{M}=\mathcal{H}^{(1,0)}+\left(\frac{\alpha}{\pi}\right)&\left[\mathcal{J}^{(1,0)}\mathcal{H}^{(1,0)}+\mathcal{S}^{(1,0)}\mathcal{H}^{(1,0)}+\mathcal{H}^{(2,0)}\right]\nonumber\\
+\left(\frac{\alpha_s}{\pi}\right)&\left[\mathcal{J}^{(0,1)}\mathcal{H}^{(1,0)}+\mathcal{S}^{(0,1)}\mathcal{H}^{(1,0)}+\mathcal{H}^{(1,1)}\right]\nonumber\\
+\left(\frac{\alpha_s}{\pi}\right)\left(\frac{\alpha}{\pi}\right)&\left[\left(\mathcal{J}^{(1,1)}+\mathcal{J}^{(0,1)}\mathcal{J}^{(1,0)}+\mathcal{J}^{(1,0)}\mathcal{J}^{(0,1)}+\mathcal{S}^{(1,1)}\right)\mathcal{H}^{(1,0)}
\right.\nonumber\\
&\left.+\right(\mathcal{J}^{(1,0)}+\mathcal{S}^{(1,0)}\left)\mathcal{H}^{(1,1)}+
\right(\mathcal{J}^{(0,1)}+\mathcal{S}^{(0,1)}\left)\mathcal{H}^{(2,0)}+\mathcal{H}^{(2,1)}\right].
\end{align}

Using these definitions, we were able to reproduce the result of this reference. In the calculation one can define $4$ gauge invariant subsets by considering the charges $Q_l$ and $Q_q$ and the quantum number $C_F$. As an example, consider the contributions proportional to $Q_q^2Q_l^2$. Diagramatically, the finite contribution to the hard scattering matrix element is then defined by:
\begin{align}
    &\Graph[.6]{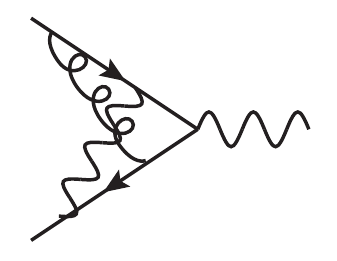}-\left(\mathcal{J}_{Q_q^2}^{(1,0)}+\mathcal{S}_{Q_q^2}^{(1,0)}\right)\left(\Graph[.6]{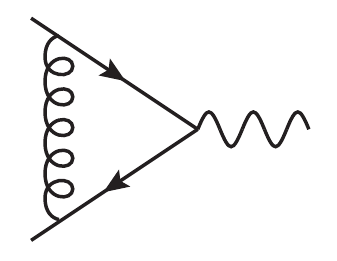}-\left(\mathcal{J}_{C_F}^{(0,1)}+\mathcal{S}_{C_F}^{(0,1)}\right)\Graph[.4]{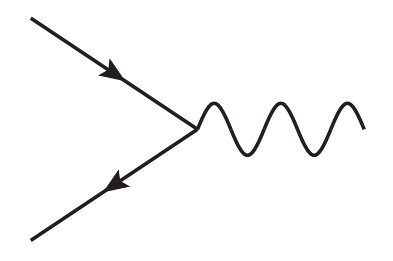}\right)\nonumber\\&
    -\left(\mathcal{J}_{C_F}^{(0,1)}+\mathcal{S}_{C_F}^{(0,1)}\right)\left(\Graph[.6]{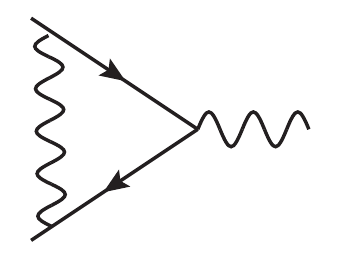}-\left(\mathcal{J}_{Q_q^2}^{(1,0)}+\mathcal{S}_{Q_q^2}^{(1,0)}\right)\Graph[.6]{Tree}\right)\nonumber\\
    &-\left(\mathcal{J}_{C_F Q_q^2}^{(1,1)}+\mathcal{S}_{C_F Q_q^2}^{(1,1)}\right)\Graph[.6]{Tree}= \text{finite}.\label{ExampleIR}
     \end{align}
     Note, that in Eq.~\eqref{ExampleIR} each diagram is meant as a sum over all diagrams of the corresponding gauge class, e.g.\ the first diagram is a representative of all two-loop diagrams proportional to $C_F Q_q^2Q_l^2$. The subscripts of $\mathcal{J}$ and $\mathcal{S}$ are to be understood as instructions to select only specific subsets of the jet- and soft-functions, proportional to $Q_q^2$, $C_F$, or $C_F Q_q^2$. 

\subsection{Extension to the full EW sector}
To extend the calculation to the EW sector, one encounters several complications. The master integrals become much more difficult due to the additional mass scale of the gauge bosons. Furthermore, one has to deal with $\gamma_5$ in dimensional regularization and a non-trivial overlap of IR and UV divergences. 

The method to calculate the new master integrals with an additional mass scale was
already discussed in Sec.~\ref{masterInt}. For $\gamma_5$ we used Kreimer's scheme \cite{Korner:1991sx,KreimerPHD}, in which one gives up the cyclicity of the trace in order to maintain an anticommuting $\gamma_5$. Since traces over $\gamma_5$ matrices lead to four-dimensional Levi-Civita tensors which are contracted with $d$ dimensional loop momenta, we employ a Passarino-Veltman tensor decomposition. For a general $R_\xi$  gauge we decompose tensor integrals with up to rank $10$, which we achieve using finite field methods \cite{Kauers:2008zz,vonManteuffel:2014ixa} for the involved linear algebra.

For the UV renormalization we calculate the wave function counterterms for all particles in the on-shell scheme. Note, that the only genuine two-loop counter terms come from fermion self-energies and terms proportional to $n_f$, the latter of which we ignore here.
We renormalize the strong and electroweak couplings and the particle masses in the $\overline{MS}$ scheme, but keep the setup flexible enough to facilitate a convenient transition to other possible renormalization schemes.
After subtraction of all UV divergences one is left with IR divergences only. Since these originate from the photon or the gluon, one expects that the same jet and soft functions can be used as in the case of QCD-QED, cf.\ Eq.~\eqref{SoftJet}.

As an example,  consider the gluon plus Z corrections to the $q\overline{q}\gamma$ vertex.
In this case we encounter one- and two-loop
counterterms for the wave-function renormalization:
\begin{align}
    &\Graph[.6]{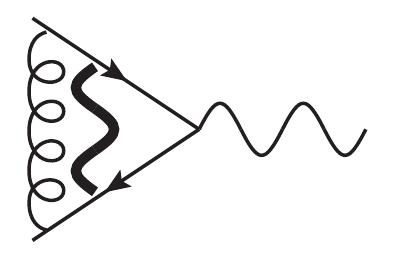} + \Graph[.6]{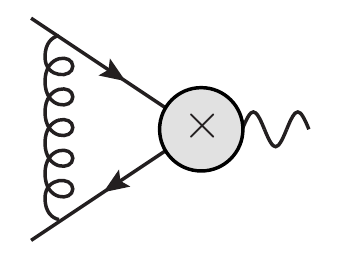}+ \Graph[.6]{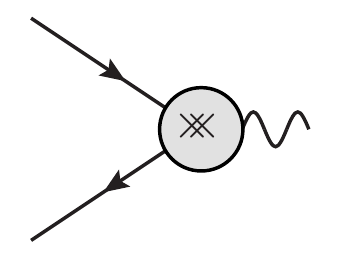}-\left(\mathcal{J}_{C_F}^{(0,1)}+\mathcal{S}_{C_F}^{(0,1)}\right)\times\nonumber\\
    &\left(\Graph[.6]{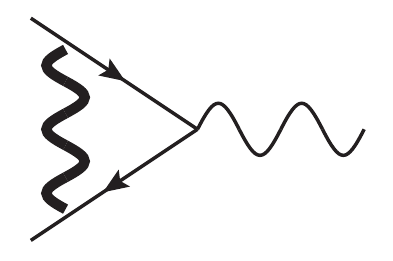}+\Graph[.6]{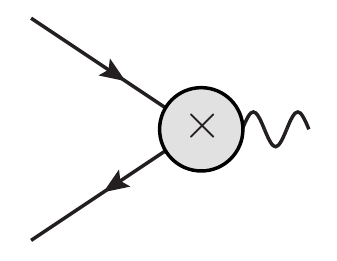}\right)= \text{finite}.\label{ExVertex}
\end{align}
In Eq.~\eqref{ExVertex} the thick lines denote the $Z$ boson and, as before, each diagram is a representative of a set of gauge invariant diagrams.
In a similar way, we arrive at the finite remainder for the other contributions to the amplitude.

\section{Conclusion and outlook}

We calculated all mixed QCD-EW virtual two-loop corrections to the Drell-Yan production of a charged lepton pair.
The master integrals have symbol letters, which depend on an unrationalizable square root.
For the first time, we integrated such differential equations in terms of multiple polylogarithms with algebraic arguments.
Our analytic solution allows for a fast numerical evaluation suitable for phenomenological applications.
We checked that the infrared structure of the two-loop amplitude matches the structure predicted by the soft and jet-functions available in the literature from the calculation for the two-loop QCD-QED corrections.

Our two-loop amplitudes provide a crucial building block for future calculations of very precise cross sections and distributions for the Drell-Yan process,
including the high-invariant mass region.
The new methods we developed to calculate the master integrals motivate an investigation of whether other Feynman integrals with a similar analytic structure may be treated analogously.

\section*{Acknowledgments}
AvM was supported in part by the National Science Foundation under Grant No.\ 1719863. MH was supported in part by the German Research Foundation (DFG), through the Collaborative Research Center, Project ID 204404729, SFB 1044, and the Cluster of Excellence PRISMA$^+$, Project ID 39083149, EXC 2118/1.

\end{document}